\documentclass[useAMS,usenatbib,usegraphicx,onecolumn]{mn}
\usepackage{graphicx}
\usepackage{amsmath}
\usepackage{amsfonts}
\usepackage{sw20mnra}

\input{tcilatex}
\begin{document}

\title{Coupled spin models for magnetic variation of planets and stars\\
}
\author{A. Nakamichi$^{\ast 1}$, H Mouri$^{\ast 2}$, D. Schmitt$^{\ast 3}$ ,
A. Ferriz-Mas$^{\ast 4}$, J. Wicht$^{\ast 3}$, \and and M. Morikawa$^{\ast
5} $ \\
$^{\ast 1}$Koyama Astronomical Observatory, Kyoto Sangyo University, \\
Motoyama, Kamigamo, Kita-ku, Kyoto 603-8555, Japan\\
$^{\ast 2}$Meteorological Research Institute, \\
1-1 Nagamine, Tsukuba 305-0052, Japan \\
$^{\ast 3}$2Max-Planck-Institut fur Sonnensystemforschung, 37191 \\
Katlenburg-Lindau, Germany\\
$^{\ast 4}$3Departamento de Fisica Aplicada, Universidad de Vigo, \\
32004 Orense, Spain\\
$^{\ast 5}$Department of Physics, Ochanomizu University,\\
2-1-1 Otsuka, Bunkyo, Tokyo 112-8610, Japan}
\date{\today \\
corresponding author Morikawa: hiro@phys.ocha.ac.jp}
\maketitle

\begin{abstract}
Geomagnetism is characterized by intermittent polarity reversals and rapid
fluctuations. We have recently proposed a coupled macro-spin model to
describe these dynamics based on the idea that the whole dynamo mechanism is
described by the coherent interactions of many small dynamo elements. In
this paper, we further develop this idea and construct a minimal model for
magnetic variations. This simple model naturally yields many of the observed
features of geomagnetism: its time evolution, the power spectrum, the
frequency distribution of stable polarity periods, etc. This model has
coexistent two phases; i.e. the cluster phase which determines the global
dipole magnetic moment and the expanded phase which gives random perpetual
perturbations that yield intermittent polarity flip of the dipole moment.
This model can also describe the synchronization of the spin oscillation.
This corresponds to the case of sun and the model well describes the
quasi-regular cycles of the solar magnetism. Furthermore, by analyzing the
relevant terms of MHD equation based on our model, we have obtained a
scaling relation for the magnetism for planets, satellites, sun, and stars.
Comparing it with various observations, we can estimate the scale of the
macro-spins.
\end{abstract}

\shortauthor{Nakamichi, Mouri. Schmitt, Ferriz-Mas, Wicht and Morikawa}
\keyphrases{dynamo, MHD, Sun: dynamo, planets and satellites: magnetic fields, }

\section{Introduction}

Geomagnetism is still one of the unsolved fundamental problems of the Earth
since William Gilbert discovered the magnetized earth (Gilbert 1600). Not
only the very existence of the dipole magnetism, Motonori Matsuyama
discovered the polarity reversal events (Matsuyama 1927) and vivid dynamics
of geomagnetism. The qualitative nature of this polarity flip was explained
by a simple disk model by Tsuneji Rikitake (Rikitake 1957). Recent numerical
simulations of magnetohydrodynamics (MHD) for geomagnetism (Roberts \&
Glatzmaier 2000), (Kono \& Roberts 2002) seem to describe such dynamics.
Inspired by these simulations, we recently studied to clarify the physics
behind this remarkable polarity reversals proposing a simple model (domino
model) (Mori, Schmitt, Ferriz-Mas et al. 2011) composed from many
macro-spins which are interacting with each other. This \textsl{coupled spin
model }is based on the idea that the whole dynamo mechanism is described by
global interactions of many small dynamo elements (macro-spins). This model
naturally yields many of the observed features of geomagnetism: its time
evolution, the power spectrum, the frequency distribution of stable polarity
periods, etc. (Mori et al. 2011). In the case of the earth, the dynamo
element, that a macro-spin describes, is considered to be the Taylor cell in
the iron fluid core produced and supported by the Coriolis force.

In this paper, we would like to develop the above idea by elaborating the
domino model in a form that is \textsl{minimal and general}.

In order to construct the \textsl{minimal }model, we introduce a \textsl{%
long-range coupling }of macro-spins, i.e. all the spins interact with each
other with the same amplitude of interactions. This is essentially different
from the previous domino model in which only neighboring spins interact with
each other. It turns out that this long-range coupling model is also
effective to describe the observed features of geomagnetism as the domino
model.

In order to show the \textsl{generality }of the model, we consider it in
wider contexts. The spin model, so far, is at most a phenomenological model
to describe geomagnetism. Therefore it needs some \textsl{supplement from
the basic MHD equations }for better descriptions. On the other hand,
magnetic fields associated with celestial objects are quite common in the
universe; many planets, satellites, and the stars including our Sun.
Although the magnetic fields of these objects have variety that is quite
different from the geomagnetism, all the magnetic fields are thought to be
generated and supported by the robust dynamo mechanism. Therefore we try to
fit our model, supplemented with MHD equations, in \textsl{other celestial
objects }and see how they are consistent with each other.

The long-range coupled spin model can also describe the \textsl{%
synchronization }physics as well if we simply change the parameter value in
the model. This property is particularly useful to describe the
quasi-periodic variation of solar magnetic field. Furthermore the long-range
coupling in our model itself can produce sufficient chaoticity and
randomness without introducing explicit random force as in the domino model.
Thus the long-range coupling spin model has rich physics and generality.

We start our study by the basic description for geomagnetism. The \textsl{%
fluid iron }in the center of the earth is now considered to be the place
where geomagnetism is created (Inglis 1981). The conductive fluid motion and
its interaction with electromagnetic fields should control the dynamics of
geomagnetism through dynamo mechanism. They are described by the \textsl{%
magneto-hydrodynamics} (MHD)\ for incompressible ($\func{div}\vec{v}=0$)
fluid, 
\begin{equation}
\begin{array}{c}
\rho \left( {\frac{\partial \vec{v}}{\partial t}+}\left( {\vec{v}\cdot \vec{%
\nabla}}\right) {\vec{v}}\right) =-\vec{\nabla}\left( {p-\frac{1}{2}\rho
\left\vert {\vec{\Omega}\times \vec{r}}\right\vert ^{2}}\right) +\rho \nu
\Delta \vec{v} \\ 
+\rho \vec{g}-2\rho \vec{\Omega}\times \vec{v}+\vec{j}\times \vec{B},%
\end{array}
\label{eqforv}
\end{equation}%
\begin{equation}
\frac{\partial T}{\partial t}+\vec{v}\cdot \vec{\nabla}T=\kappa _{T}\Delta
T+\varepsilon _{T},  \label{eqforT}
\end{equation}%
\begin{equation}
\frac{\partial \vec{B}}{\partial t}=\vec{\nabla}\times \left( {\vec{v}\times 
\vec{B}}\right) +\eta \Delta \vec{B},  \label{eqforB}
\end{equation}%
\begin{equation}
\vec{j}=\frac{1}{\mu _{0}}\vec{\nabla}\times \vec{B},  \label{jB}
\end{equation}%
with appropriate boundary conditions (Roberts \& Glatzmaier 2000). The
dynamics of fluid velocity ${\vec{v}}$\ in Eq.(\ref{eqforv}) is governed by
the nonlinearity $\left( {\vec{v}\cdot \vec{\nabla}}\right) {\vec{v},}$ the
pressure $p,$ the centrifugal force potential $-(\rho /2){\left\vert {\vec{%
\Omega}\times \vec{r}}\right\vert ^{2}}${, }gravity $\rho \vec{g}$, and the
Lorentz force $\vec{j}\times \vec{B}$ where $\vec{j}$ is the electromagnetic
current and $\rho $ is the mass density. The Coriolis force $-2\rho \vec{%
\Omega}\times \vec{v}$ yields the vorticity and convective pattern of the
fluid. The magnetic field $\vec{B}$ is amplified or reduced by the flow $%
\vec{v}$ as described by Eq.(\ref{eqforB}). The dissipations ($\nu \Delta 
\vec{v},\kappa _{T}\Delta T,\eta \Delta \vec{B}$) and the energy injections
(from the heat generation $\varepsilon _{T},$ and possibly the rotational
driving force)\ should balance with each other in the stationary state. Main
non-dimensional quantities which characterize the above set of equations,
and their typical values in the iron fluid core in the earth, are as follows:%
\begin{equation}
{c}R_{e}=\left\vert \frac{{\vec{v}\cdot \nabla \vec{v}}}{\nu \Delta \vec{v}}%
\right\vert =O\left( 10^{8}\right) ,  \label{Re}
\end{equation}%
\begin{equation}
\sqrt{T_{a}}=\left\vert \frac{-2\vec{\Omega}\times \vec{v}}{\nu \Delta \vec{v%
}}\right\vert =O\left( 10^{14}\right) ,  \label{Ta}
\end{equation}%
\begin{equation}
\sqrt{R_{a}}=\left\vert \frac{\vec{g}}{\sqrt{\nu \kappa _{T}}\Delta \vec{v}}%
\right\vert =O\left( 10^{8}\right) ,  \label{Ra}
\end{equation}%
\begin{equation}
R_{m}=\left\vert \frac{\nabla \times \left( {\vec{v}\times \vec{B}}\right) }{%
\eta \Delta \vec{B}}\right\vert =O\left( 10^{2}\right) .  \label{Rm}
\end{equation}%
These values are important to estimate the scale of dynamo elements in later
sections.

Straightforward approach based on these highly complicated equations
requires sophisticated numerical calculations. Such analysis have recently
been developed and the generation, maintenance and even reversals of
geomagnetism have been obtained (Kida \& Kitauchi 1998),(Roberts \&
Glatzmaier 2000),(Kono \& Roberts 2002). These works have provided us with
valuable knowledge to develop fundamental understanding of geomagnetism.
However the parameter range in numerical calculations is still far from the
above realistic values. Therefore it would be very favorable if our coupled
spin model is complementary to the MHD calculations and if it describes some
basic features of dynamo mechanism. These are the aim of this paper.

The plan of this paper is as follows. In \textsl{section 2}, we describe our
idea of the coupled dynamo elements which plays the central role in the
present paper. We emphasize the necessary condition for the elements in
order to clarify what the macro-spin represents in our model. Readers who
immediately want to see the model can skip this section. In \textsl{section 3%
}, we briefly review our previous study on the domino model (short-range
coupled spin model (SCS)), which is a first realization of the idea of the
coupled dynamo elements to describe geomagnetism. In \textsl{section 4}, we
introduce the generalized long-range coupled-spin model (LCS) in which
macro-spins have long range interactions. Then we apply this model to
geomagnetism and successfully describe many of the characteristic features
of geomagnetism. In \textsl{section 5}, we study the physical relevance of
LCS model showing the similarity of it to the mean field model (HMF) and
Kuramoto model. Readers who immediately want to see further applications of
the model can skip this section. In \textsl{section 6}, we apply our model,
supplemented with MHD equations, to other planets and satellites in order to
estimate the scale of macro-spins. In \textsl{section 7}, we apply the same
model, with different parameter values, to the solar magnetism and
successfully describe the quasi-periodic dynamics of the solar magnetism.
The last \textsl{section 8 }is devoted to summarize our work, and speculate
possible generalizations of our idea of coupled dynamo elements.

\section{Coupled dynamo elements}

We now describe the idea of coupled dynamo elements (Mori et al. 2011) which
plays the fundamental role in this paper. The central equation to describe
the dynamo mechanism in MHD is Eq.(\ref{eqforB}), in which the flow $\vec{v}$%
\ is determined by Eq.(\ref{eqforv}). This is practically a very complicated
set of equations to describe geomagnetism and the dynamo mechanism though we
can extract some basic insight from it.

Geomagnetism is characterized by the coexistence of many time scales, from
million years to thousand years, with clear power-law power spectrum. The
dynamo mechanism is ubiquitous in the sense that most of the celestial
objects have magnetism generated from this mechanism. The ubiquitousness
implies the basic dynamo mechanism is simple and the many time scales with
power-law power spectrum imply the dynamo system is composed from many
components. Therefore we expect that \textsl{the whole dynamo system may be
composed of many elements, each of which has minimal dynamo function, that
interact with other elements in a simple form}. It is often happen that a
simple interaction yields very complicated structures.

Then what is the element which has the minimal dynamo function in the case
for geomagnetism? We notice that the above Eq.(\ref{eqforB}) is similar to
the general equation for vorticity $\vec{\omega}\equiv rot\vec{v},$ 
\begin{equation}
\frac{\partial \vec{\omega}}{\partial t}=rot\left( \vec{v}\times \vec{\omega}%
\right) +\nu \Delta \vec{\omega},
\end{equation}%
which is derived by taking the rotation of Eq.(\ref{eqforv}), neglecting the
Lorentz force term. This fact reveals the apparent duality of the magnetic
field $\vec{B}$ and the vorticity $\vec{\omega}$. The iron fluid core is
estimated to be highly turbulent from Eq.(\ref{Re}) and is full of
vorticities. The coherent vorticity $\vec{\omega}$\ is generated from the
convection and Coriolis force. These effects seem to be significant from the
values in Eqs.(\ref{Ta}-\ref{Ra}). Therefore the iron fluid is full of
vorticity, and possibly of magnetic fields.

It is clear that the Coriolis force dominates in the iron fluid as in Eq.(%
\ref{Ta}). In general, if the Coriolis force dominates in almost the
stationary flow, then Eq.(\ref{eqforv}) becomes $\vec{\Omega}\times \vec{v}=-%
\vec{\nabla}p/\left( 2\rho \right) .$ By taking the rotation of this form,
we have $\left( \vec{\Omega}\cdot \vec{\nabla}\right) \vec{v}=0$. Therefore
the flow has a tendency to be two-dimensional perpendicular to the
rotational axes $\vec{\Omega}$, \textsl{i.e.} the Taylor-Proudman theorem.
Therefore this theorem suggests that a convection of iron fluid forms
convective columns (usually called Taylor cell) parallel to the rotation
axes. The scale of the Taylor cell will be determined by geometry and
various parameters in Eqs.(\ref{Re}-\ref{Ra}). Thus coherent vorticity $\vec{%
\omega}$\ is naturally expected in the earth core.

However the vorticity $\vec{\omega}$\ is not sufficient to produce magnetism 
$\vec{B}$ (Tsinober 2007). This is because the circular vorticity itself is
not endowed with energy production. Any deviation from the complete circular
motion would be necessary for the production of magnetic energy. We would
like to emphasize that the inward-winding vorticity is essential for dynamo
mechanism. Suppose that we prepare a typical inward-winding solution for the
Navier-Stokes equation, the Burgers vortex, whose velocity field is
expressed in the cylindrical coordinate $\{r,\theta ,z\}$ as

\begin{equation}
v_{\theta }\left( r\right) =\frac{2\nu \omega _{0}}{\alpha r}\left( 1-e^{-%
\frac{\alpha }{4\nu }r^{2}}\right) ,\;v_{r}\left( r\right) =-\frac{1}{2}%
\alpha r,\;v_{z}\left( z\right) =\alpha z  \label{bergers}
\end{equation}%
with some constants $v,\omega _{0},\alpha $, and $\alpha >0$ for
inward-winding and $\alpha <0$ for outward-winding. Then putting the above
Eq.(\ref{bergers}) into Eq.(\ref{eqforB}) yields, in the Cartesian
coordinate,

\begin{equation}
{c}\frac{\partial B_{3}}{\partial t}=\alpha B_{3}-(v_{r}\cos \theta
-v_{\theta }\sin \theta )\frac{\partial B_{3}}{\partial x^{1}}-(v_{r}\sin
\theta +v_{\theta }\cos \theta )\frac{\partial B_{3}}{\partial x^{2}}+\text{%
dissipative terms, }
\end{equation}%
for the magnetic field $B_{3}\left( t,x^{1},x^{2},x^{3}\right) $ parallel to
the symmetry axes of the vorticity. Only the first term in RHS can
definitely enhance $B_{3}$ if $\alpha >0.$ The inward-winding flow squeezes
the magnetic force lines against the positive pressure from $B_{3}$.

Summarizing all the above, the iron fluid in the earth core may allow 
\textsl{multiple vortex columns parallel to the earth rotation axes and
their interaction may cooperatively yield the whole geomagnetism, if the
columns are inward-winding}.

On the other hand in the super computer simulations, such column structure
is commonly formed (Kida \& Kitauchi 1998) and robustly persists. These
columns are called Taylor cell(TC). A close look into TC in the numerical
data, reveals that TC are either inward-winding (\textsl{i.e.} high-pressure
due to the compression) or outward-winding (\textsl{i.e.} low-pressure due
to the spread). They are called, respectively, anti-cyclone and cyclone.
That is the anti-cyclone counter rotates to the earth and the cyclone
co-rotates. The Coriolis force associated with the earth rotation makes
these difference and the smooth flow may favor the alternating configuration
of cyclone and anti-cyclone\footnote{%
However the clear Taylor column structure is not fully verifyed in numerical
calculations of MHD. For example, the vorteces are in the form of sheets
instead of columns \ (Kageyama \& Miyagoshi 2008). In either cases, the idea
of the coupled dynamo elements would be effective if such sheets or columns
have dynamo function.}. According to the above argument, only the
anti-cyclone, which has inward-winding flow, can produce magnetic fields
(Kageyama \& Sato 1997). In the real situations in the earth, the flow is
highly turbulent and non-linear on top of such TC structures reflecting the
large Reynolds number Eq.(\ref{Re}). Complicated magnetic fields in the
outer core are captured by inward-winding TC and aligned with the rotation
axes, and then compressed to enhance the poloidal magnetic fields. These
non-linear flow, as well as huge electric current, yields the interaction
among TC, either short-range or long-range.

Each dynamo element of inward-winding vorticity, anti-cyclone, accompanies
generated magnetic fields which can be characterized by a direction and a
strength in the first approximation. Therefore we characterize each dynamo
element as a \textsl{vector }quantity $\vec{s}$. This vector represents the
direction of the magnetic fields associated with TC and not the direction of
TC itself. Since each element is supposed to have the minimum dynamo
function, this vector can be identified as a macroscopic spin or magnetic
moment. Then the whole system should be a set of such macro-spins $\vec{s}%
_{i}$, which is located on each site $i$ in the iron fluid core. Although
the interactions between the macro-spins should be very complicated, the
potential energy of it must be a \textsl{scaler }or the vector inner product
of the spins $\lambda \vec{s}_{i}\cdot \vec{s}_{j}$. This simplicity may
guarantee the robustness and universality of the geomagnetism.

However we cannot determine the amplitude of the coupling parameter $\lambda 
$ by a simple argument. It depends on at least the iron flow pattern,
electric currents, and configuration of cyclone and anti-cyclone. On the
other hand we can guess the signature of $\lambda $. Electric currents are
smoothly winding each TCs, and cyclone and anti-cyclone align alternately.
Therefore the smooth electric current yields the winding direction is the
same for all anti-cyclones (and opposite direction for all cyclones). This
suggests the magnetic fields of anti-cyclones themselves (and also cyclones
themselves) have a tendency to align to the same direction. This suggests the%
\textsl{\ negative value for the coupling strength:} $\lambda <0$\footnote{%
A naive consideration that the macro-spin \ behaves as a bar magnet leads to
the opposite signature: positive $\lambda $.}, like in the ferromagnetism%
\footnote{%
Negative $\lambda $ in ferromagnetism has the quantum origin, which has
nothing to do with our macro-spin model.}. Then the effective `Curie point`
of these macro-spin system can be high enough and therefore, the whole
system may actually show `ferromagnetism`. This makes a quite contrast to
the case of the (micro) spins of the iron in the earth; no ferromagnetism is
produced in the core of temperature $\approx 6000$K which well exceeds the
Curie point about $\approx 1000$K.

We have emphasized the necessary condition for the macro-spin $\vec{s}_{i}$\
to be relevant and their interaction although the construction of the
macro-spin $\vec{s}_{i}$ is not our aim in this paper. We would like to
study the inter-relation between many of such elements $\vec{s}_{i}$\ which
yields various non-trivial dynamics.

Before our study, there have been some works (Mazaud \& Laj 1989), (Seki \&
Ito 1993), (Dias, Franco, \& Papa 2008) in which the spin models with the
interaction $\lambda \vec{s}_{i}\cdot \vec{s}_{j}$ were studied to reveal
the polarity flip dynamics of geomagnetism. They are based on the
two-dimensional Ising model and have found the polarity flip near the
critical temperature. Our aim to study geomagnetism in this paper is not
simply to extend the similarity with phase transitions but to study wider
point of view such as \textsl{phase coexistence} and \textsl{synchronization}%
.

There are two types of coupling between the spins according to the range of
the interaction. If the fluid motion is responsible to the coupling, then
only the neighboring spins interact with each other and yield short-range
coupled-spin (\textsl{SCS}) system. On the other hand, if the electric
current or magnetic fields are responsible to the coupling, then all the
spins interact with the same strength and yields long-range coupled-spin
system (\textsl{LCS}). These are the two extreme ends of the series of
models with finite interaction rage. We have already studied SCS model in
our paper (Mori 2011), which successfully describes many characteristic
features of geomagnetism as we now review in the next section before we
study LCS model is the subsequent sections.

\section{Review of short-range coupled spin (SCS) model}

We now briefly review our previous model for geomagnetic dynamics (Mori et
al. 2011), in which the concept of coupled dynamo elements is realized as
short-range interacting macro-spins (SCS model). Our intention was to
establish a simple model which realizes various basic properties of
geomagnetic observations based on the idea that the coupled dynamics of
dynamo elements. We assume that each element has dynamo function to produce
element magnetism putting the very origin of the dynamo mechanism aside.

We first introduce the SCS model, which is described by the Lagrangian $%
L=K-V,$ where%
\begin{equation}
K\equiv \frac{1}{2}\sum_{i=1}^{N}\left( \frac{d\vec{s}_{i}}{dt}\right) ^{2}=%
\frac{1}{2}\sum_{i=1}^{N}\dot{\theta}_{i}^{2},  \label{K}
\end{equation}%
\begin{equation}
V\equiv \mu \sum_{i=1}^{N}\left( \vec{\Omega}\cdot \vec{s}_{i}\right)
^{2}+\lambda \sum_{i=1}^{N}\vec{s}_{i}\cdot \vec{s}_{i+1},  \label{V}
\end{equation}%
and by appropriate dissipation and fluctuation terms. The spins are assumed
to be located on a circle with equal separation (therefore $\vec{s}_{N+1}=%
\vec{s}_{1}$). This circle is set on the plane which includes the equator
and fully immersed inside the fluid core region of the earth. The spin $\vec{%
s}_{i}$ at the cite $i$ may be represented by a single angle parameter $%
\theta _{i}$ as $\vec{s}_{i}=\left( \cos \theta _{i},\sin \theta _{i}\right) 
$. We measure the angle $\theta _{i}$\ from the angular velocity vector $%
\vec{\Omega}$ \footnote{%
The configuration is shown in Fig.(\ref{fig6-1}). The angular velocity $\vec{%
\Omega}$ points to the top.}. This plane the spin $\vec{s}_{i}$ moves is set
perpendicular to the radius toward the cite $i$. The full evolution equation
then becomes the stochastic differential equation (Langevin equation),%
\begin{equation}
\ddot{\theta}_{i}=-\frac{\partial V}{\partial \theta _{i}}-\kappa \dot{\theta%
}_{i}+\xi _{i},  \label{EOMLCS}
\end{equation}%
for $1\leq i\leq N$. The parameter $\kappa $ is the dissipation coefficient.
The fluctuation force $\xi _{i}$ is assumed to be Gaussian white noise as
usual, characterized by the correlation:%
\begin{equation}
<\xi _{i}\left( t\right) \xi _{j}\left( t^{\prime }\right) >=2\varepsilon
\delta _{ij}\delta \left( t-t^{\prime }\right) .
\end{equation}%
Then we have, so far, four parameters in our model $\mu ,\lambda ,\kappa
,\varepsilon $. The parameter $\mu $ represents the tendency for each spin
to be parallel to the rotational axes. Larger value of $\mu $ means the
dominance of the Coriolis force over the other effects and the
Taylor-Proudman theorem better holds. The parameter $\lambda $ measures the
strength of interaction between the element spins. It reflects the nature of
the fluid flow and electric current across the neighboring spins. The
parameter $\kappa $ represents the whole energy dissipation $\nu ,\eta ,$
and $\varepsilon _{T}$ in Eqs.(\ref{eqforv},\ref{eqforB}, \ref{eqforT}). The
parameter $\varepsilon $ represents the inhomogeneous heat generation from
the inner iron core and the random perturbation from the other element
spins. We also have to specify the number $N$ of spins. It may depend on the
geometry of the fluid core region of the earth in which cyclones and
anti-cyclones are formed.

We can save the number of parameters if we focus on the dynamics in long
time scale and neglect the inertial term $\ddot{\theta}_{i}$. Then the
reduced equation of motion becomes ($1\leq i\leq N$),%
\begin{equation}
\dot{\theta}_{i}=-\frac{\partial V}{\partial \theta _{i}}+\xi _{i}\text{,}
\label{redLCS}
\end{equation}%
which now has three parameters $\mu ,\lambda ,\varepsilon $. The full
equation Eq.(\ref{EOMLCS}) is solved in (Mori et al. 2011). The neglect of
the inertial term is justified by the fact that both the methods yield
almost the same results.

Each dynamo element represented by the spin $\vec{s}_{i}$ naturally
corresponds to a magnetic dipole located at the site $i$\ of the dynamo
element. We are interested in the order parameter $\vec{M}\left( t\right) $
of the set of equations Eq.(\ref{redLCS}) defined by 
\begin{equation}
\vec{M}\left( t\right) \equiv \frac{1}{N}\sum_{i=1}^{N}\vec{s}_{i}\left(
t\right) .  \label{M}
\end{equation}%
This quantity $\vec{M}\left( t\right) $($\in \lbrack -1,1]$) is a good
indicator of the whole magnetic field when we analyze the history of the
geomagnetism though it is not the magnetic field itself. It is also possible
to describe the realistic configuration of the magnetic field on the earth
by using the set of $\vec{s}_{i}\left( t\right) $, ($1\leq i\leq N$). As we
have explained before, the magnetic moments $\vec{s}_{i}$ ($1\leq i\leq N$)
are located on the ring with equal separation, on the surface which includes
the equator. Then the whole magnetic field $\vec{B}$ becomes 
\begin{equation}
\vec{B}\left( \vec{y}\right) =\sum_{i=1}^{N}\frac{3\left( \vec{s}_{i}\cdot 
\vec{n}_{i}\right) \vec{n}_{i}-\vec{s}_{i}}{\left\vert \vec{y}-\vec{x}%
_{\left( i\right) }\right\vert ^{3}},  \label{B}
\end{equation}%
where each spin $\vec{s}_{i}$ is located at the position $\vec{x}_{\left(
i\right) }$, and $\vec{n}_{i}$ is the unit vector $\vec{n}_{i}\equiv \left( 
\vec{y}-\vec{x}_{\left( i\right) }\right) /\left\vert \vec{y}-\vec{x}%
_{\left( i\right) }\right\vert $.

This simple model naturally describes many of the observed properties of
geomagnetism including the intermittent polarity reversal, power-law power
spectrum, chron interval distribution, etc. The results are summarized in
the paper (Mori et al. 2011).

\section{Long-range coupled spin model}

The short-range coupled spin (SCS) model reviewed in the last section was
simple and effective to describe the statistical properties of geomagnetism.
We now study a slightly modified model introducing the long-range
interaction coupled spin (LCS) model. The simplest one would be described by
the same Lagrangian $L=K-V$ with Eq.(\ref{K}, \ref{V}) but the interaction
of macro spins is long ranged \footnote{%
The inclusion of the facter $N$ in the denominator in the coupling term is
simply from a technical reason so that all the terms in $V$ formally becomes
additive.},

\begin{equation}
V\equiv \mu \sum_{i=1}^{N}\left( \vec{\Omega}\cdot \vec{s}_{i}\right) ^{2}+%
\frac{\lambda }{2N}\sum_{i<j}^{N}\vec{s}_{i}\cdot \vec{s}_{j}.
\label{VinGCS}
\end{equation}%
According to the naive extension of SCS, the full evolution equation seems
to become ($1\leq i\leq N$),%
\begin{equation}
\ddot{\theta}_{i}=-\frac{\partial V}{\partial \theta _{i}}-\kappa \dot{\theta%
}_{i}+\xi _{i}.
\end{equation}%
as in the previous section, introducing the fluctuation force $\xi _{i}$ and
the dissipation term $-\kappa \dot{\theta}_{i}$. However, thanks to the
long-range coupling, these \textsl{fluctuation-dissipation terms are not
necessary in LCS }model. The long-range coupling by itself can yield
sufficient chaoticity which is necessary to describe the intermittent
polarity flip. It is still possible to reserve the fluctuation-dissipation
terms, which does not alter the characteristic feature of the system.
Therefore we would like to choose the simpler description which has no
fluctuation-dissipation terms. Then the system becomes conservative system
in which the total energy is conserved.

The evolution equation becomes simple ($1\leq i\leq N$),%
\begin{equation}
\ddot{\theta}_{i}=-\frac{\partial V}{\partial \theta _{i}},  \label{EOMGCS}
\end{equation}%
which has only two parameters $\mu $ and $\lambda $ through the potential $V$
as Eq.(\ref{VinGCS}). The order parameter and the magnetic fields are
described by the same expressions Eqs.(\ref{M}, \ref{B}) as in the SCS model.

We next report the numerical results of LCS model with comparison with
observations. We have solved the set of equations for $N=9$ spins which are
located on a circle. We have set parameters of the model as $\mu =-1,\lambda
=-1.8$, all of which is chosen to be of order one and we did not perform
systematic fine tuning of them. Further we set the initial condition so that
all the spins are rest and their direction angles are equally separated
within the width of angle $0.65\times 2\pi $. The reason that we did not
distribute the initial spin angles in the full range $[0,2\pi ]$ is because
we had to choose low energy to make the coexistence possible of the cluster
and expanded spins, as we will explain in the next section.

Time evolution of the total magnetic moment projected on the rotation axes $%
M\equiv \vec{\Omega}\cdot \vec{M}\left( t\right) $ is shown in Fig.(\ref%
{fig4-1}) top. The end time $6\times 10^{4}$ is chosen so that the total
number of polarity-reversal becomes several hundreds as is observed number $%
338$ in the history of geomagnetism within $1.6\times 10^{8}$ year in the
past. The total number of the polarity reversal is $199$ in this
calculation. Therefore the unit calculation time ($\equiv time$) corresponds
to about $3.1\times 10^{3}$ year. We have obtained the following results.

\begin{figure}[tbp]
\begin{center}
\includegraphics[width=12cm]{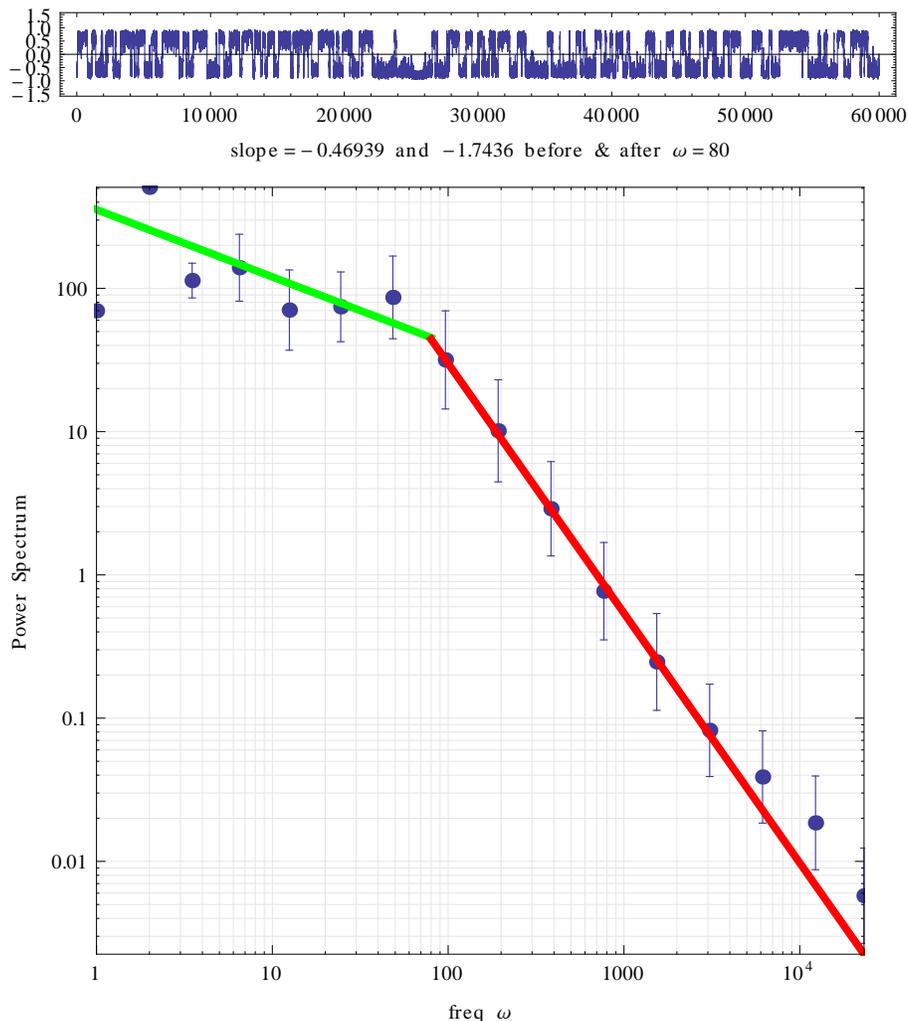} \vspace{-0.2in}
\end{center}
\caption{ (top) A typical time series of $M\left( t\right) \equiv \vec{\Omega%
}\cdot \vec{M}\left( t\right) $ \ in the numerical calculation of LCS model.
The positive and negative values in the vertical axis correspond,
respectively, to the normal and the reversed polarity. (botom) The power
spectrum of this time series. The data is sucessfully fitted by the the two
power laws with the indeces -0.5 for low frequency and -1.7 for high
frequency regions. }
\label{fig4-1}
\end{figure}

\begin{enumerate}
\item unit calculation time: $time=$ $3.1\times 10^{3}$ year\ 

\item average flipping time: $1.5\times 10^{3}$ yeas. (observation $%
(2-3)\times 10^{3}$year)

\item average chron time: $\left( 0.3-6.2\right) $ $\times 10^{6}$ year.
(observation $\left( 0.1-6\right) \times 10^{6}$ year.)

\item superchron: $4.7\times 10^{6}$ year (observation $3.5\times 10^{7}$
year )

The power spectrum of the time series $M\left( t\right) $ is shown in Fig.(%
\ref{fig4-1}) bottom.

\item the power index: $-1.7$ and $-0.47$, respectively, for right and left
of the knee. (observation: index $-1.8$ in high frequency regime $\omega
>\omega _{\ast }=3.7\times 10^{-4}$, and is $-0.67$ in the low frequency
regime $\omega _{\ast }>\omega $. )

\item the location of the knee: $1.9\times 10^{6}$ year (observation: $1.19$%
My)

These values are not much changed for the projected time series $%
signature(M\left( t\right) )$ because the original time series $M\left(
t\right) $\ is already almost the same as the projected shape, relatively
long steady chrons and sharp reversal.

\item The frequency of chron intervals is calculated. the power index: $-1.4$
(observation: $-1.52$)

\item Next we pick up a typical chron and calculated the power spectrum
within this period.

the power index: $-2.3$ for $4.7\times 10^{5}$year \ (ODP: $-2.2$ for $%
\omega >2.8\times 10^{-3}$)
\end{enumerate}

\begin{figure}[tbp]
\begin{center}
\includegraphics[width=12cm]{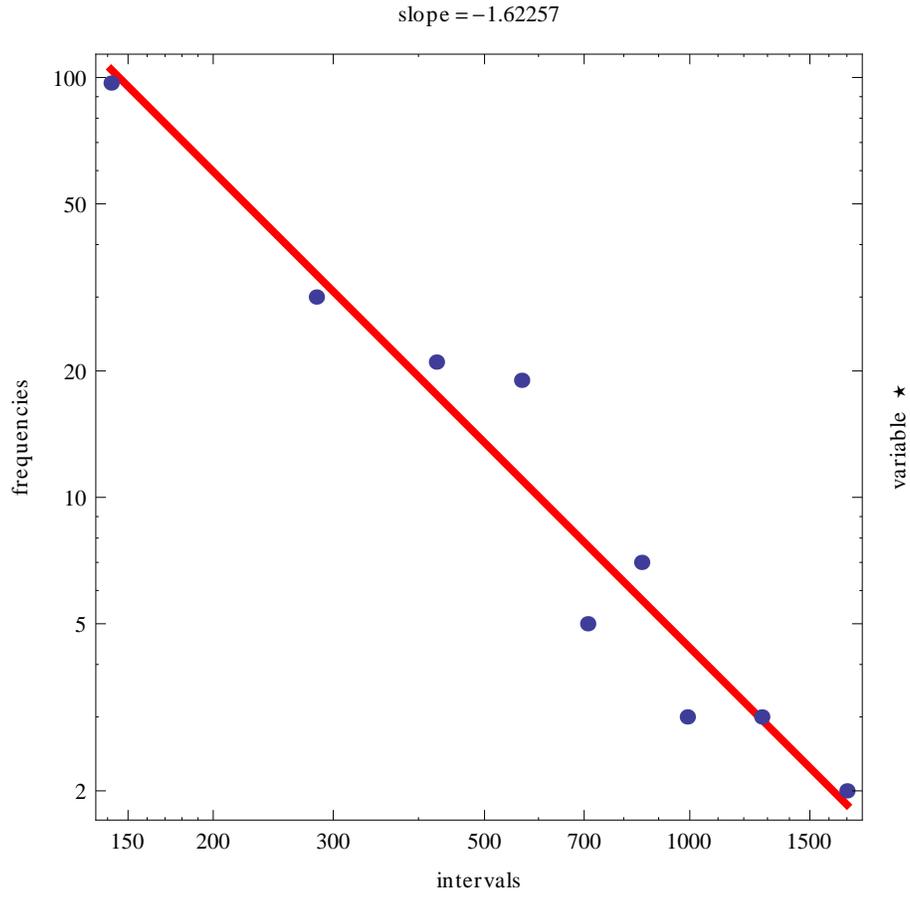} \vspace{-0.2in}
\end{center}
\caption{ The frequency of the intervals of fixed polarity. The distribution
is very roughly fitted by a power law with the index -1.6. }
\label{fig4-2}
\end{figure}

\begin{figure}[tbp]
\begin{center}
\includegraphics[width=12cm]{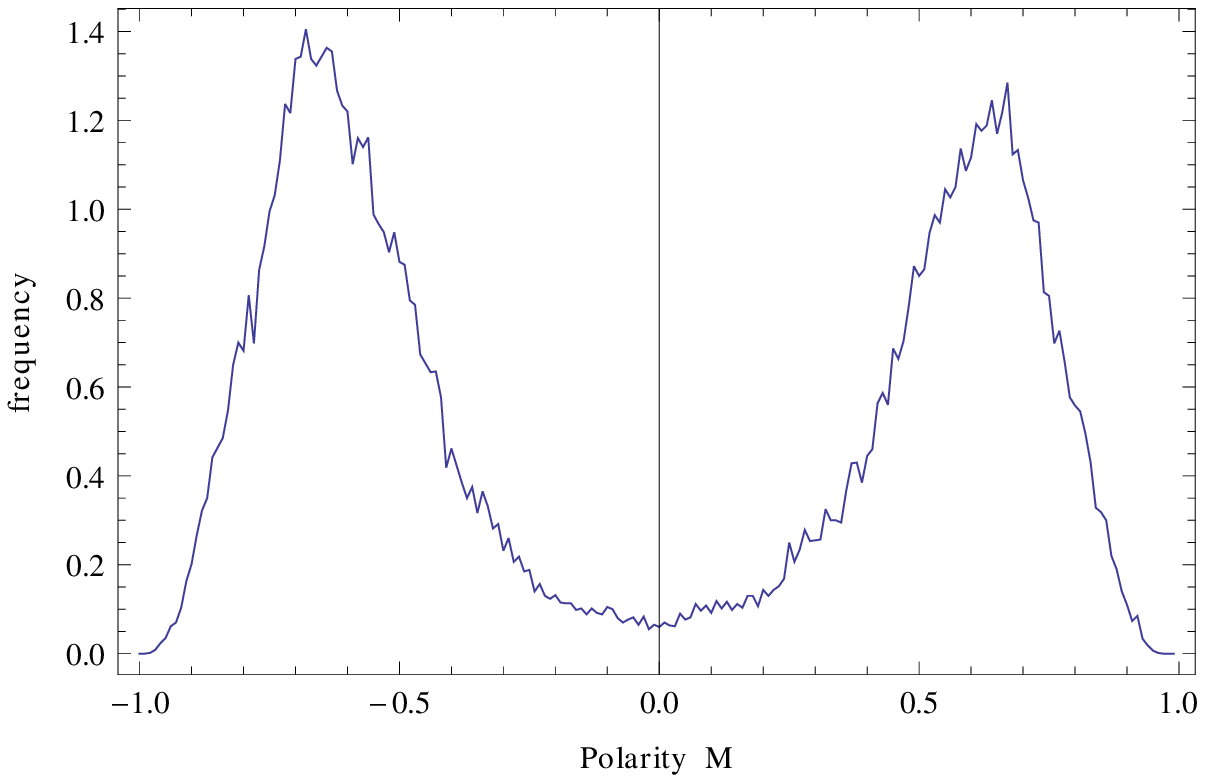} \vspace{-0.2in}
\end{center}
\caption{ The total frequency distribution of $M\left( t\right) $. The
left-right symmetry, or the symmetry in normal-reversed polarites, is not
respected. This indicates the model has strong fluctuation in the long time
variation. This accord with the ever increasing power spectrum toward low
frequency region Fig.(\protect\ref{fig4-1}). }
\label{fig4-3}
\end{figure}

The apparent asymmetry in Fig.(\ref{fig4-3}) is not caused by the initial
condition but suggests long range temporal fluctuations in our model. This
seems to be consistent with the ever increasing power spectrum in the
low-frequency direction in Fig.(\ref{fig4-1}).

\begin{figure}[tbp]
\begin{center}
\includegraphics[width=12cm]{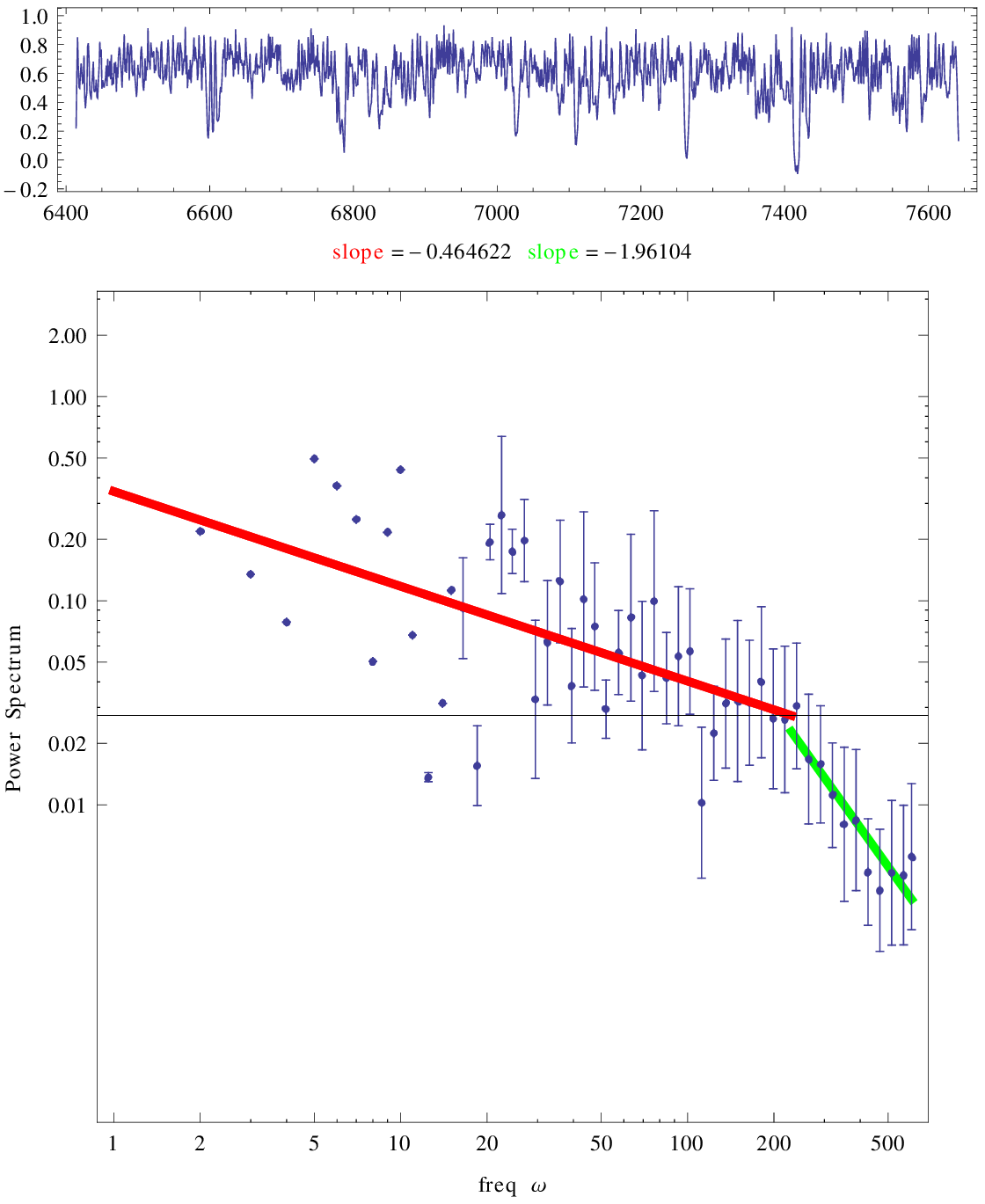} \vspace{-0.2in}
\end{center}
\caption{ (top) A typical time series of $M\left( t\right) \equiv \vec{\Omega%
}\cdot \vec{M}\left( t\right) $ \ in some chron. The valuea are mostly
negative but perpetually fluctuating strongly. (botom) The power spectrum of
this time series. The data is sucessfully fitted by the the two power laws
with the indeces -0.46 for low grequency and -2.0 for high frequency
regions. }
\label{fig4-4}
\end{figure}

\begin{figure}[tbp]
\begin{center}
\includegraphics[width=12cm]{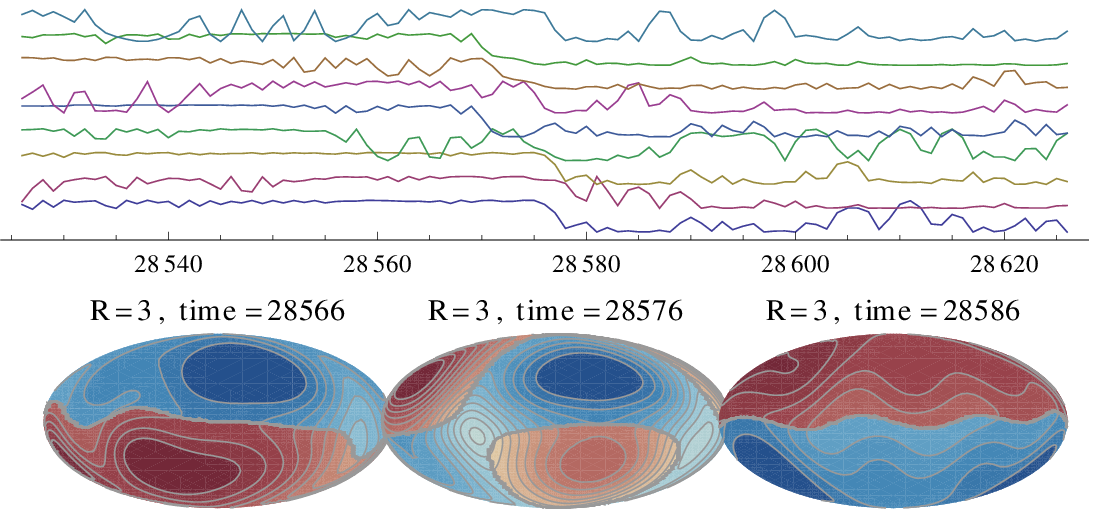} \vspace{-0.2in}
\end{center}
\caption{ (top) Individual spin motion around a typical polarity flip at
time 31264. Vertical axes represents $\vec{\Omega}\cdot \vec{s}_{i}\left(
t\right) $ for each spin with shifted zero points for visibility. The
first spin from top was rapidly rotating before the flip but the nineth
spin took over this role after the flip. These perpetually
rotating spins form the gas phase. See the next section for detail. (bottom)
The countour/density graph for horizontal component of the magnetic field
Eq.(\protect\ref{B}) on the earth surface around this polarity flipping era.
Each ellipse represents the entire earth surface drown in Mollweide mapping,
before, right in the middle, and after the flip from left to right. The
radius of the spin ring is assumed to be one-third of the Earth radius. }
\label{fig4-5}
\end{figure}

The results of LCS model for geomagnetism is almost the same as our previous
study on SCS model and are not so far from the observational values. Thus
the range of interaction is shown to be irrelevant for geomagnetic
characteristics. In the SCS model, even the inertial term was not necessary
to yield the similar results.

\section{General properties of LCS model}

It is interesting to notice that this LCS system, we explored in the last
section, is a slight extension of the \textsl{Hamiltonian Mean Field model }%
(HMF (Antoni \& Ruffo 1995), (Campa, Dauxois , \& Ruffo 2009) by adding the
potential term $\mu \sum_{i=1}^{N}\left( \vec{\Omega}\cdot \vec{s}%
_{i}\right) ^{2}$. HMF model a good tool to describes phase transitions and
statistical fluctuations of a deterministic system composed from many
elements. Slightly modified version of HMF model was applied to describe the
self gravitating systems(Sota, Iguchi, Morikawa, Tatekawa, \& Maeda 2001).
However this HMF model itself cannot describe the polarity flip of
geomagnetism because of the lack of the preferred angle. The polarity flip
is made possible simply by introducing the above potential term. In this
sense our LCS model is the \textsl{minimal model }to describe geomagnetism.

Acute readers might have some concern that the geodynamics is a dissipative
system rather than the conservative system like HMF model. However the
difference is not relevant in our context. HMF and LCS systems have strong
chaoticity and the mean field $M$ has apparent statistical fluctuations. In
short, these systems have the relevant central degrees of freedom (\textsl{%
i.e.} mean field) and the rest (\textsl{i.e. }environmental degrees of
freedom) together. We focus on the mean field. The mean field immersed in
the rest behaves as the dissipative system. This point will be clarified in
the rest of this section.

Dynamics of both HMF and LCS models can be visualized by the interacting $N$%
-particles which are moving on a circle of unit radius as shown in Fig.(\ref%
{fig5-1}). This is a typical snapshot of the spin angles $\{\theta
_{i}\}\left( i=1...N\right) $ represented by the points on a circle. The
whole ring represents all the directions of spins $0\leq \theta <2\pi $ and
the top position of the ring is the direction of the earth rotation axes $%
\vec{\Omega},$ $\theta =0$.

\begin{figure}[tbp]
\begin{center}
\includegraphics[width=12cm]{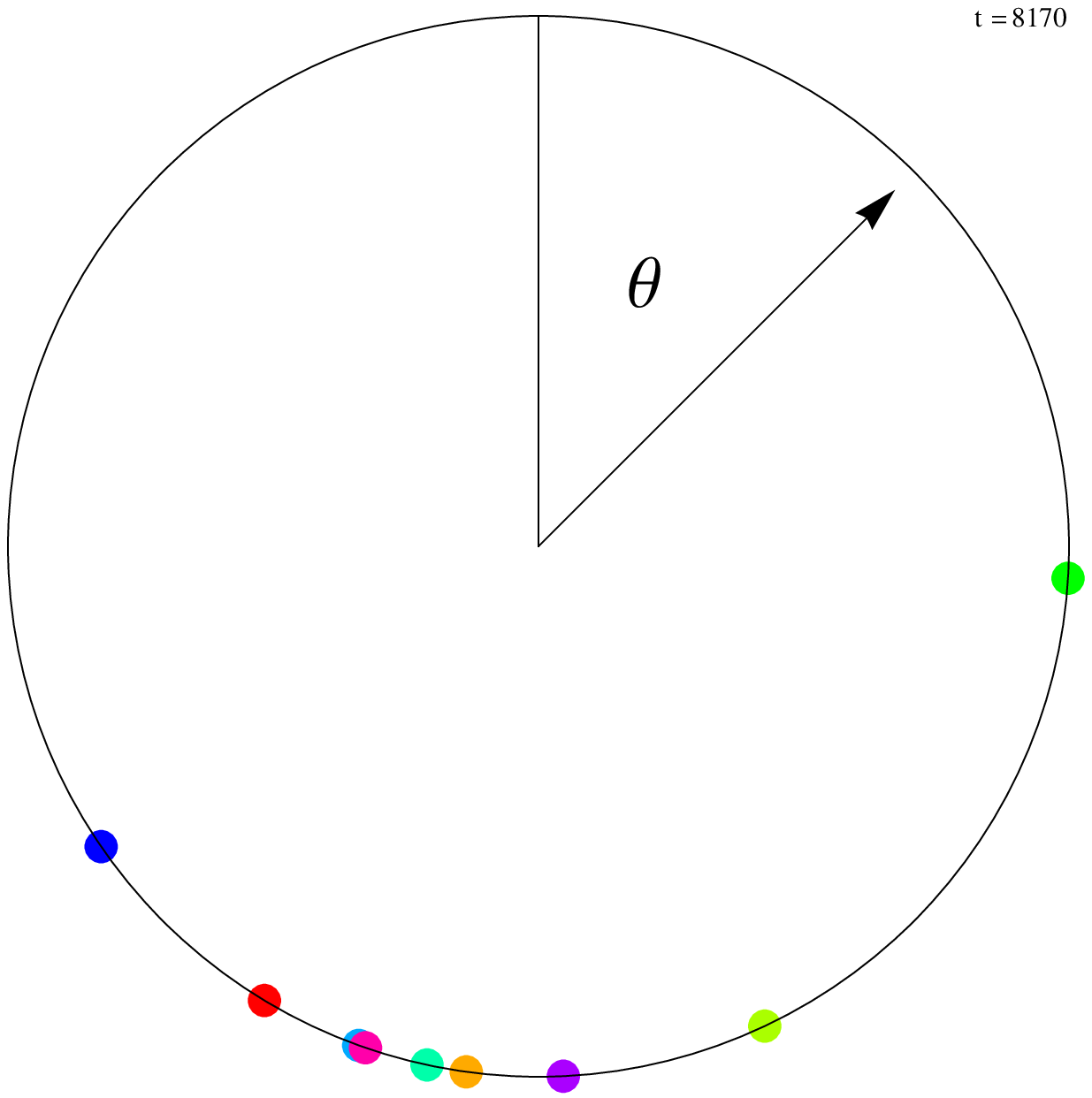} \vspace{-0.2in}
\end{center}
\caption{ A typical snapshot of the spin angles $\protect\theta _{i}\left(
i=1...N\right) $. The whole ring represents $0\leq \protect\theta <2\protect%
\pi $ and the top position of the ring is $\protect\theta =0$. The earth
rotation axes $\vec{\Omega}$ has the fixed angle $\protect\theta =0$. As is
shown the most spins are condenced to the almost fixed angle (i.e. fixed
polarity), and few rapidly rotating spins exit. The latter trigges the flip
of the condensed spins intermittently. }
\label{fig5-1}
\end{figure}

Furthermore it is remarkable that both HMF and LCS models have two classes
(phases) of elements: the spins in \textsl{expanded phase }which are almost
freely moving all the angle and the spins in \textsl{condensed phase} which
are bounded in a localized angle. There exists a critical energy in HMF
model in the limit $N\rightarrow \infty $; the condensed phase appears only
when the system energy is below this critical energy and the portion of the
condensed phase increases for lower energy. Similarly in our LCS model,
expanded and condensed phases coexist in wide rage of parameters\footnote{%
There is no sharp critical energy since our $N$ is not so large.}. We have
chosen our parameter so that the system energy is low enough to guarantee
that the both phases coexist in one system. The condensed phase yields
finite order parameter defined by Eq.(\ref{M}). Free spins in the expanded
phase perpetually disturb the condensed cluster spins and is considered to
trigger the intermittent signature change of the order parameter Eq.(\ref{M}%
). It is interesting that the expanded and condensed phases often exchange
their constituent spins when the angles of the spins coincide with each
other, \textsl{i.e.} `collide` in the above visualization.

In terms of geomagnetism, coexistence of the expanded and condensed phases
are essential; the latter yields the well defined polarity in magnetism Eq.(%
\ref{M}) and the former triggers the intermittent flip of the polarity. This
phase coexistence is naturally realized in LCS model when we choose mildly
low energy values\footnote{%
If we chose high enough energy then the dominant dipole magnetic field would
not be formed. If we chose too low energy then it would take very long time
for the dipole flip.}. Thus, this model naturally deduces that the system
energy required for intermittent polarity flip necessitates \textsl{the
coexistence of the relatively steady component and the rapidly changing
component in geomagnetism}. The former is of course from the condensed phase
and the latter from the expanded phase. Since the macro-spin has magnetic
dipole, these freely moving spins in the expanded phase yields \textsl{%
rapidly moving pairs of local magnetic spots of reverse polarity} superposed
on the average steady dipole magnetism caused by the spins in condensed
phase. These pairs also have a tendency to align on north-south direction
according to the potential form in Eq.(\ref{VinGCS}). These pairs have some
similarity with The South Atlantic Magnetic Anomaly (SAMA) observed on the
core-mantle boundary (Korte 2010).

The above interpretation that the spins in the expanded phase trigger the
flip of the global magnetic polarity suggests that a possible \textsl{%
correlation between the activity of the spins in expanded phase and the
duration of the fixed polarity}. Actually in our numerical calculations, the
expanded phase temporally disappears in the superchron, i.e. the longest
period of fixed polarity. This correlation should be further studied.
Anomalous acceleration in the movement of the local magnetic spots of
reverse polarity may predict a polarity flip in the near future.

It is also interesting to notice that this set of equations Eq.(\ref{EOMGCS}%
) is a slight extension of the \textsl{Kuramoto model for synchronization}%
(Kuramoto 2003, Acebron, Bonilla, Perez, Ritort, \& Spigler 2005) by the
change of potential: $\mu \sum_{i=1}^{N}\left( \vec{\Omega}\cdot \vec{s}%
_{i}\right) ^{2}\rightarrow -\sum_{i=1}^{N}\omega _{i}\theta _{i}$, and the
reduction of the order of time derivative: $\ddot{\theta}_{i}\rightarrow 
\dot{\theta}_{i}$. The potential $-\sum_{i=1}^{N}\omega _{i}\theta _{i}$
forces each angle $\theta _{i}$\ to rotate with the fixed frequency $\omega
_{i}$ despite the lack of the inertial term. Therefore the latter
simplification $\ddot{\theta}_{i}\rightarrow \dot{\theta}_{i}$ is possible
in Kuramoto model. Kuramoto model is a set of long-range coupled oscillators
with different frequencies. If there were no interaction between the
oscillators, then each phase of the oscillator behaves independently.
However if the long-range interaction is sufficiently strong, then the
phases of oscillators \textsl{synchronize }with each other despite the
difference in each frequencies $\omega _{i}$. Kuramoto model very generally
describes the synchronization processes of many type of oscillators and
limit cycles including chemical and biological systems.

It is remarkable that LCS model can also describe the synchronization of
element spins and yield \textsl{quasi-periodic motion of the mean field}, by
simply changing the parameters. The first term of RHS in Eq.(\ref{VinGCS})
gives independent non-linear oscillations of each spin angle. Since the
oscillation is nonlinear and the amplitudes are different from spin to spin,
the phases of the spins behave very chaotically. The second term of RHS in
Eq.(\ref{VinGCS}) gives the cosine potential between all the spin pairs ($%
\vec{s}_{i}\cdot \vec{s}_{j}=\cos \left( \theta _{i}-\theta _{j}\right) $)
and thus naturally yields the tendency for all the pairs to synchronize with
each other. Whether the whole synchronization actually takes place or not
depends on the balance between the two terms. If the pairwise interaction $%
\lambda $\ is sufficiently large in LCS model, then the synchronization
takes place as in the case of Kuramoto model. However, reflecting the strong
non-linearity of the cosine potential, the synchronized oscillation becomes
quasi-periodic in general. This potential ability of LCS model to describe
synchronization physics is especially important when we apply LCS to the
solar magnetism in later sections.

\section{Other Planets and celestial objects}

We would like to argue, in this section, to what extent the LCS\ model can
be \textsl{general} in wider contexts. The concept of the coupled dynamo
elements itself, as we have discussed in section 2, is general and is not in
principle restricted to geomagnetism. We expect the same for LCS model,
which is the minimal realization of this concept. The magnetic fields
associated with celestial objects are quite common in the universe; many
planets, satellites, and the stars including our Sun. Although the magnetic
fields of these objects have variety that is quite different from the
geomagnetism, all the magnetic fields are thought to be generated and
supported by the robust dynamo mechanism.

However the LCS model is at most a phenomenological model to describe dynamo
systems. The LCS model is simply a macroscopic description and cannot be
directly deduced from the microscopic MHD equations Eqs.(\ref{eqforv}-\ref%
{jB}), which are especially important for quantitative argument in wider
contexts. Therefore we \textsl{need to supplement LCS model with the MHD
equations}.

First we examine the \textsl{energy flow} for geomagnetism. Geomagnetic
fields would be amplified by the inward-winding motion of the convection
vortex, anti-cyclone. This inward motion is caused by the Coriolis force
induced by the earth rotation although the rotation itself cannot directly
transfer its energy to geomagnetism. On the other hand, the convective
motion of the iron fluid is supported by the thermal flow from the central
region of the earth. This flow is mainly supported by the release of the
gravitational potential by the core sinking. Thus the thermal flow
accelerates the convective motion and supports the inward-winding force to
amplify the magnetic fields through the Coriolis force.

One essential fact for this thermal energy flow is that the main dissipation
is due to the Joule heat. This is because the magnetic Prandtl number $%
P_{m}\equiv \nu /\eta $, i.e. the ratio of the time scale of magnetic
diffusion ($1/\eta $)\ and that of convective diffusion ($1/\nu $), is
actually very small: $P_{m}\approx 10^{-6}\ll 1$.

We now closely look at the basic equations Eqs.(\ref{eqforv}, \ref{eqforB})
in the above picture of energy flow. From Eq.(\ref{eqforB}), we first notice
that the steady state for the magnetic field requires the \textsl{energy
balance on average}: $\nabla \times \left( {\vec{v}\times \vec{B}}\right)
+\eta \Delta \vec{B}=0$, which reduces to the relation which is independent
of $\vec{B}$, 
\begin{equation}
v\approx \left( \gamma \mu _{0}\sigma R_{c}\right) ^{-1},  \label{v-steady}
\end{equation}%
where $R_{c}$ is the Fe-core radius and $\gamma R_{c}$ is the radius of the
Taylor cell. The existence of this special scale is essential to our model
to characterize the dynamo elements; most of the relevant process of dynamo
takes place on this scale. This small scale compared with the core size is
foreseen from the order in Eq.(\ref{Rm}). The magnetic field is amplified by
the inward winding motion of the convection vortex through the term $\nabla
\times \left( {\vec{v}\times \vec{B}}\right) $ until this steady state Eq.(%
\ref{v-steady}) is realized. On the other hand, the inward-winding motion of
flow is generated by the term $-2\rho \vec{\Omega}\times \vec{v}$ in Eq.(\ref%
{eqforv}), and this term should eventually \textsl{balances }with the
backreaction from the generated magnetic pressure through the term $\vec{j}%
\times \vec{B}$\ in the same equation. Therefore in the steady state, we
have the estimate of the typical strength of the magnetic field $B_{in}$ at
the scale $\gamma R_{c}$, the scale of the Taylor cell, as

\begin{equation}
B_{in}\approx \left( 2\rho _{0}\Omega v\mu _{0}\gamma R_{c}\right) ^{1/2}.
\label{Bin-steady}
\end{equation}%
The observed magnetic field $B_{out}$\ at the Earth surface is related with
this estimated $B_{in}$\ through the magnetic flux conservation as, 
\begin{equation}
B_{out}=\left( \frac{\gamma R_{c}}{R}\right) ^{2}NB_{in},  \label{BoutBin}
\end{equation}%
where $N$ is the number of Taylor cells in the core and $R$\ is the Earth
radius. We simply assumed that all the Taylor cell is the same scale and the
magnetic fields associated with TC are parallel to the rotation axes of the
earth for simplicity. Combining the above three equations, we have 
\begin{equation}
B_{out}=N\gamma ^{2}\left( \frac{R_{c}}{R}\right) ^{2}\left( \frac{2\rho
_{0}\Omega }{\sigma }\right) ^{1/2},  \label{Bout}
\end{equation}%
and the corresponding (virtual) magnetic moment $d$\ becomes%
\begin{equation}
d\equiv B_{out}R^{3}=N\gamma ^{2}R_{c}^{2}R\left( \frac{2\rho _{0}\Omega }{%
\sigma }\right) ^{1/2}.  \label{d}
\end{equation}

In the geomagnetic case, we have chosen $N=9$ and putting reasonable values, 
$R=6357$km, $R_{c}=3480$km, $\sigma =3\times 10^{5}\func{Si}$emens/m, $\rho
_{0}=5497$kg/m$^{3}$, and assuming $\gamma =10^{-1}$, we have $%
B_{out}=0.4\times 10^{-4}$Tesla from Eq.(\ref{Bout}), and $v=7.62\times
10^{-6}$m/$\sec $ from Eq.(\ref{v-steady}). Although the value of $B_{out}$\
is consistent with observations the value of $v$\ is only $8\%$ of the
estimated speed of the convection flow simply from the west-ward drift
motion. If this discrepancy is real, the time scale of the magnetic field
variation or of the electric current change is about ten times larger than
the convective flow speed. This point should be further studied.

\begin{figure}[tbp]
\begin{center}
\includegraphics[width=12cm]{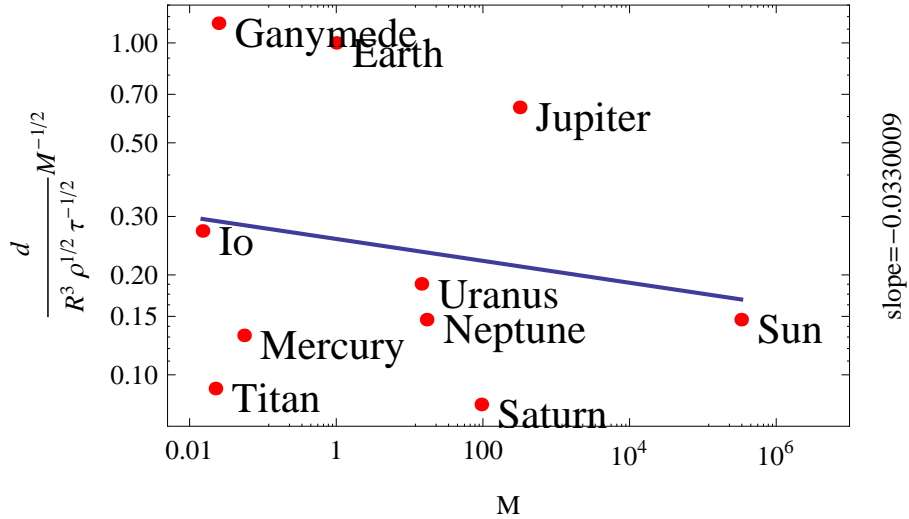} \vspace{-0.2in}
\end{center}
\caption{ The quantity $d/(R^{3}\protect\sqrt{2\protect\rho _{0}\Omega /%
\protect\sigma }M^{1/2})$ is shown in the vertical axes against the mass $M$
in the horizontal axes for various celestial objects. These quantities are
normalized by the values of earth. The data are from Stevenson (2010).
Eulopa and Callisto are excluded from the figure because their magnetic
fields are probably induced from the external fields (Stevenson 2010). The
value $4\times 10^{-4}$T is used for the solar dipole magnetic field. The
quantity $d/(R^{3}\protect\sqrt{2\protect\rho _{0}\Omega /\protect\sigma }%
M^{1/2})$ turns out to be an almost constant within a factor $10$ for the
mass range over 8 digits. This fact suggests the scaling $N\protect\gamma %
^{2}\propto M^{1/2}$, which determins the number and the radius of the
dynamo element in general. }
\label{fig6-1}
\end{figure}

The estimate of the effective magnetic moment in Eq.(\ref{d}) is general
since it is deduced simply from the basic equations (\ref{eqforv}, \ref%
{eqforB}) based on the LCS model. The expression Eq.(\ref{d}) is consistent
with the general claim in the literature (Christensen2010), (Stevenson 2010)
that the planetary magnetic fields are essentially determined from the
factor $\sqrt{2\rho _{0}\Omega /\sigma }$. We can step further and estimate
the unknown factor $N\gamma ^{2}$ in Eq.(\ref{d}). By using various data for
celestial objects, but assuming $\sigma $ is a constant, we can draw a graph
for $d$ against $R^{3}\sqrt{2\rho _{0}\Omega /\sigma }$\footnote{%
We simply set $R_{c}\approx R$.}. This turns out to show a power law
relation with the index about 3/4\footnote{%
This is essentially the same as the well known relation between the magnetic
moment $d$ and the angular moment $J$\ for celestial objects $d\leq
cJ^{\beta }$ where $c=1.9\times 10^{-12}($Meter$^{2}/$Ampere)$^{0.93},$ $%
\beta =0.93$.}: $d\propto \left( R^{3}\sqrt{2\rho _{0}\Omega /\sigma }%
\right) ^{3/4}$, though this is not what we needed.\ Simply motivated by
this scaling relation, we can draw a similar graph for $d/(R^{3}\sqrt{2\rho
_{0}\Omega /\sigma }M^{1/2})$ against $M$\ in Fig.(\ref{fig6-1}). This
relation turns out to be almost a constant within a factor $10$ for the
whole mass range of 8-digits as shown in Fig.(\ref{fig6-1}).

The factor $M^{1/2}$ is important for this constancy\footnote{%
Any other extensive variable powered by appropriate index whould yield the
constancy to some extent. However this factor $M^{1/2}$ yields the smallest
variation of the data as far as we have checked. When we choose $M$ for the
extensive variable, the best index was $0.53$ which is $1/2$ within an error.%
}. We can deduce the scaling formula, from this phenomenological relation,
as 
\begin{equation}
N\gamma ^{2}\propto M^{1/2}.  \label{taylorcellscaling}
\end{equation}%
In other words, the scaling Eq.(\ref{d}) holds if we choose Eq.(\ref%
{taylorcellscaling}). This phenomenological scaling is helpful to reveal the
number of dynamo elements. For example, if we suppose the parameter $\gamma $
is almost a constant for various objects, and $N=9$\ for Earth, we have $%
N=9\times (M_{\odot }/M_{\text{earth}})^{1/2}\approx 5\times 10^{3}$ for the
sun. However, we must keep in mind that the planets, satellites and stars
are very different with each other and the structure of dynamo mechanism is
different (Stanley 2010) despite the above generality.

Keeping the above caution in mind, it is interesting to try the unified view
of planetary magnetism within our model. In the context of the LCS model,
observed large dipole tilt more than $50^{o}$ and the quadrupole moment of
magnetic fields in \textsl{Uranus and Neptune }are not exceptional. The
magnetic field described by LCS model naturally becomes irregular and
asymmetric when occasional large excursion takes place. We may observing
these irregular and asymmetric fields in the present Uranus and Neptune.
Moreover in the context of LCS model, almost axis symmetric magnetic field
of \textsl{Saturn }does not conflict with the anti-dynamo theorem, which
inhibits dynamo function in \ the axisymmetic steady configuration. This is
because individual element dynamo yields magnetic field independently from
others and easily violates axis symmetry of the whole system. The magnetic
field in LCS model is rapidly changing and steady condition is also violated.

\section{Application of LCS model to the solar magnetism}

As we have seen in the previous section, a simple argument on the MHD
equations based on the LCS model can yield a general scaling law for planets
and satellites. This was useful to estimate the scale of dynamo elements. In
this section, we would like to explore this generality of the LCS model by
applying it to the solar magnetic dynamics.

LCS model does not specify the macro-spin or dynamo element. The macro-spin
can express any subsystem which has minimal dynamo property. Typically the
element is the inward-winding vorticity. In case of our sun, the macro-spin
may be the convecting vortex deep inside the solar surface. It may happen
that this region is turbulent and the hierarchy of vortices yield a network
of many dynamo elements. The very end of this hierarchy may appear on the
entire surface of the sun, possibly as the supergranulations. Recently the
pattern of the supergranulation is observed by using the local
helioseismology (Nagashima 2010). On top of these supergranulation, strong
horizontal magnetic field is observed everywhere on the surface (Ishikawa
2008). We speculate that these vortices associated with dynamo function may
fill the entire convecting region inside the sun.

\begin{figure}[tbp]
\begin{center}
\includegraphics[width=12cm]{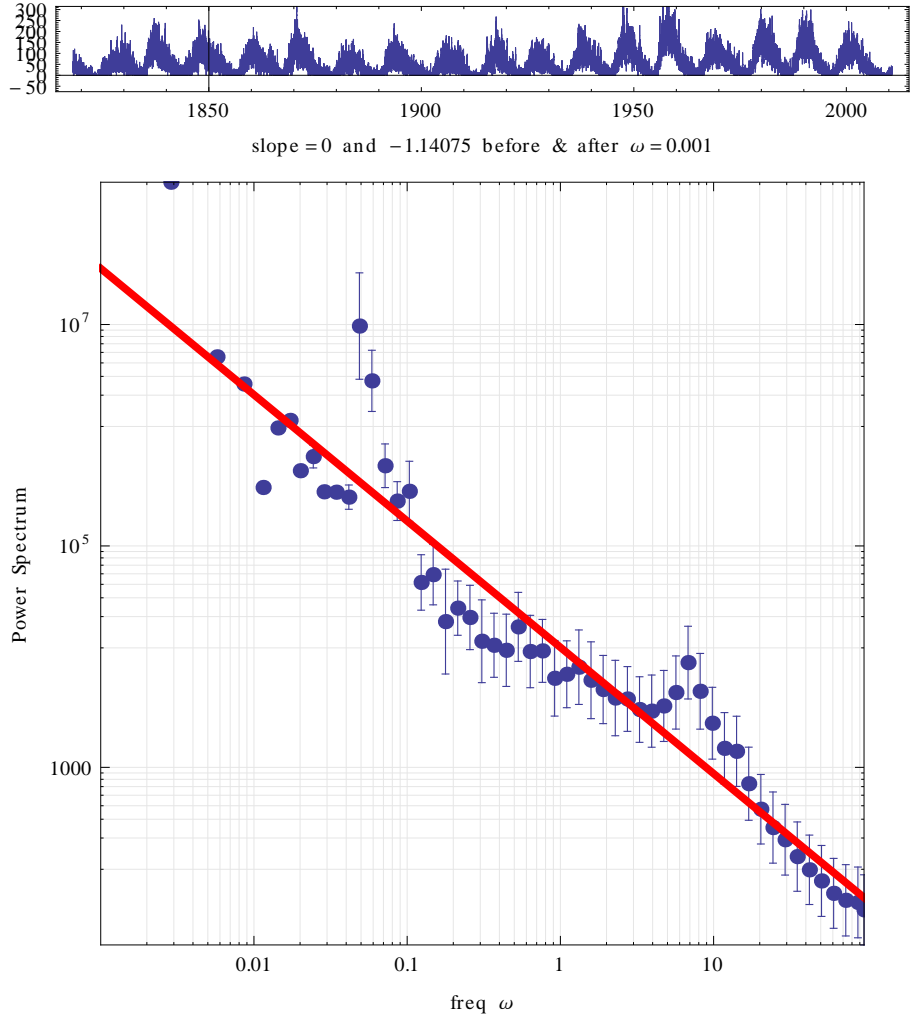} \vspace{-0.2in}
\end{center}
\caption{ (top) Dayly number of sunspots are shown in the vertical axes
against the time in year in the horizontal axes. (botom) The power spectrum
of this time series. The data is sucessfully fitted by a power law with the
index -1.1 on top of the two peaks at $\protect\omega \approx 0.05$ and $%
\protect\omega \approx 7$. The former peak corresponds to the 11 year solar
activity cycle and the latter peak fictitiously appears corresponds to a
month the solar rotation period. }
\label{fig7-1}
\end{figure}

The solar magnetism changes its polarity quasi-regularly with the period
about 22 years, which synchronizes with the solar activity of period 11
years. The solar activity can be roughly measured simply by counting the
sunspots (SIDC 2011). We show this daily count profile in Fig. (\ref{fig7-1}%
) top, during the past two hundred years. The power spectrum, shown in Fig. (%
\ref{fig7-1}) bottom, reveals the clear periodicity of about 11 years, as
well as the period of about a month. The latter period is an artifact
reflecting the systematic sunspot change on the solar surface which rotates
with this period. What is interesting for us is the scaling property with
index $-1.1$, near to $-1$, on top of the two peaks associated with the
above periods. One-over-f noise or the pink noise, \textsl{i.e.} its power
spectrum scales with index $-1$, has been observed everywhere in nature.
(For a comprehensive list of references on 1/f noise, see (Li 2011)).
Similar 1/f noise in the monthly averages of the sunspot numbers has been
reported in the high frequency ($f>\left( 11\text{year}\right) ^{-1}$)
regions (Franchiotti, Sciutto, Garcia, \& Hojvat 2004).

It is remarkable that the LCS model shows quasi-periodic polarity flip,
which resembles the solar magnetism, as a result of synchronization of many
spins. For the synchronization process to set in, we need large number of
spins ($N\gg 1$) and low potential barrier ($\left\vert \mu \right\vert \ll
1 $). Then the spins in LCS model can naturally synchronize to yield
quasi-periodic motion like in the Kuramoto model.

\begin{figure}[tbp]
\begin{center}
\includegraphics[width=12cm]{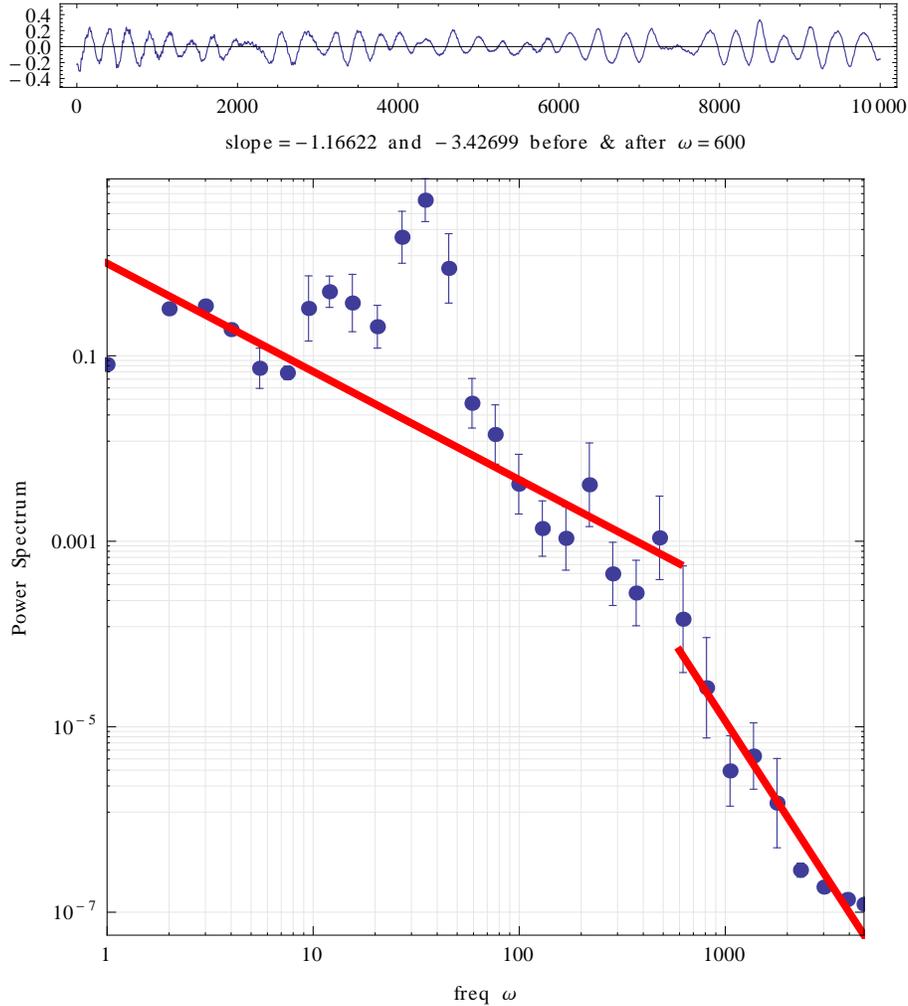} \vspace{-0.2in}
\end{center}
\caption{ (top) A typical time series of $M\left( t\right) \equiv \vec{\Omega%
}\cdot \vec{M}\left( t\right) $ \ in the numerical calculation with the
parameter $N=101,$ $\protect\mu =-0.01,$ $\protect\lambda =-2$, and the
initial condition is chosed so that all the spin angles are equally
separated within the 70\% of the whole angle $0\leq \protect\theta <2\protect%
\pi $. Each spin has the interaction with nearest $10$ spins. The
quai-periodic oscillation is apparent. This oscillation appears as a result
of synchronization of many spins. (botom) The power spectrum of this time
series. There is a peak at $\protect\omega \approx 30$ reflecting the above
quasi-periodicity on top of the rough power law behavior with index $-1.2$
in low frequency region. These behaviors accord with the solar observation
in the previous figure. }
\label{fig7-2}
\end{figure}

For example in a numerical demonstrations, we set $N=101,$ $\mu =-0.01,$ $%
\lambda =-2$, and obtained the result as shown in Fig.(\ref{fig7-2}). The
top figure shows the time evolution of the order parameter $M\left( t\right)
\equiv \vec{\Omega}\cdot \vec{M}\left( t\right) $ defined in Eq.(\ref{M}).
Quasi-periodicity in its evolution is apparent. The amplitude irregularly
varies and even becomes almost flat in some periods. The bottom figure shows
its power spectrum. Beside the apparent peak corresponding to the
quasi-periodicity, one-over-f noise, actually the power $-0.85$, is observed
in low frequency region. However this result needs to be refined before
serious discussions. For example, the number of spins $N=101$ is apparently
too small compared with the previous estimate $N\approx 5\times 10^{3}$ in
the last section.

If we suppose that the solar magnetic activity is linked with the number of
sunspots, then the above results suggest that the LCS model can reproduce
the observed solar activity especially \textsl{the quasi-periodicity as well
as 1/f like power law in the low frequency region in its power spectrum}.

\section{Summary and Discussions}

We have developed a long-range coupled macro-spin LCS model for
geomagnetism, solar magnetism, and others. We have seen this model is 
\textsl{minimal and general}.

In \textsl{section 2}, we developed the idea of the coupled dynamo elements
for general dynamo mechanism in magnetohydrodynamical systems such as
geomagnetism. The element should have inward-winding vorticity to amplify
the magnetic fields. The we introduced a macro-spin representing such
element. In \textsl{section 3}, we have reviewed the short-range coupled
spin (SCS)\ model which successfully describes geomagnetism. In \textsl{%
section 4}, we introduced the long-range coupled-spin (LCS) model which also
successfully describes geomagnetism. Thus the spin models are shown to be
useful to describe geomagnetism in the context of statistical fluctuations.

In \textsl{section 5}, we studied the physical relevance of LCS model. First
of all in this model, \textsl{two phases of spins can coexist}: The expanded
phase in which the spins can rapidly move all the directions almost freely
and the condensed phase in which the spin directions are bounded into narrow
region to form the dominant dipole moment which is almost steady. The spins
in expanded phase perpetually "collide" (see the figure \ref{fig5-1}) with
the condensed phases and trigger its intermittent polarity reversal. On the
other hand LCS model shows quasi-periodicity as well as 1/f noise on top of
it. This is the \textsl{synchronization }of the constituent spins. In both
cases, the system shows power law behavior reflecting the long range
interactions. The phase coexistence is essential for geomagnetism and the
synchronization for solar magnetism. These study taught us how the general
models such as Hamiltonian mean field (HMF) model and Kuramoto model, if
slightly modified, can describe the universality and variety of general
dynamo mechanism.

In \textsl{section 6}, we applied LCS model, supplemented with MHD
equations, to other planets and satellites. Finding a useful scaling low, we
could estimate the scale and the total number of the dynamo element, or
macro-spin, in terms of the mass $M$ of the body. In \textsl{section 7}, we
applied LCS model, with different parameter values, to the solar magnetism
and successfully describe the quasi-periodic oscillation of the solar
magnetism on top of the 1/f noise property in low frequency regions. Thus we
have actually demonstrated the universality and variety of LCS model.

Although the LCS model can describe most of the geomagnetic observations, it
cannot be deduced from the basic MHD equations. It is also possible to
generalize LCS model to endow each spin with dynamo function. This is
realized for example by setting the homopolar generator (Faraday disk) for
each dynamo element, $(1\leq i\leq N)$, and introduce appropriate
interactions between them. This is an extension of the original Rikitake
model (Rikitake 1957).

If we introduce nearest neighbor interaction, we obtain the short-range
coupled Rikitake model (SCR), for the currents $\left\{ J_{i}\right\}
_{1\leq i\leq N}$\ and rotation angles $\left\{ \omega _{i}\right\} _{1\leq
i\leq N}$,%
\begin{equation*}
\frac{dJ_{i}}{dt}=-J_{i}+\omega _{i}J_{i+1},
\end{equation*}%
\begin{equation}
\frac{d\omega _{i}}{dt}=c-\lambda J_{i}J_{i+1}.  \label{LCR}
\end{equation}
We can obtain the occasional polarity flip and power law power spectrum also
in this model. However even $N=8$ case, the time evolution of SCR model
inherits chaotic spiky profile from the original Rikitake model. The
long-range coupled version of Rikitake model (LCR)\ is also possible,%
\begin{equation*}
\frac{dJ_{i}}{dt}=-J_{i}+\omega _{i}\bar{J},
\end{equation*}%
\begin{equation}
\frac{d\omega _{i}}{dt}=c-\lambda J_{i}\bar{J},  \label{GCR}
\end{equation}
where $\bar{J}\equiv \sum_{i=1}^{N}J_{i}/N$ is the mean field. In this case
we obtain a regular and spiky oscillation due to the strong synchronization
of the elements. We have not yet performed a full parameter study. Further
extension of such models would be possible and they will be worth extensive
research.

\end{document}